\documentclass{IEEEtran}
\onecolumn
\usepackage{multirow}
\usepackage{subfigure}
\usepackage{graphicx}
\usepackage{longtable}
\usepackage{courier}
\usepackage{float}
\usepackage{amsmath,amssymb}

\usepackage[in]{fullpage}

\usepackage{draftwatermark}
\SetWatermarkLightness{0.9}

\title{Distributed Sensing and Transmission of \\ Sporadic 
Random Samples \\ in a Multiple-Access Channel}

\begin{document}
\author{Ay\c{s}e \"{U}nsal, Raymond Knopp\\
Mobile Communications Department, Eurecom, Sophia Antipolis, France\\
{\small{\texttt{ayse.unsal@eurecom.fr,
raymond.knopp@eurecom.fr}}}}
\maketitle
\thispagestyle{plain}
\begin{abstract}
  \footnote{This paper was presented [in part] at ISIT 2013, IEEE International Symposium on Information Theory, July 7-12, 2013, Istanbul, Turkey.}
This work considers distributed sensing and transmission of sporadic random samples. Lower bounds are derived for the reconstruction error 
of a single normally or uniformly-distributed finite-dimensional vector imperfectly measured by a network of sensors and transmitted with finite energy to a common receiver 
via an additive white Gaussian noise asynchronous multiple-access channel. Transmission makes use of a perfect causal feedback link to the encoder 
connected to each sensor. A retransmission protocol inspired by the classical scheme in \cite{Yamamoto79} applied to the transmission of 
single and bi-variate analog samples analyzed in \cite{EURECOM+3798} and \cite{EURECOM+3898} is extended to the more general network 
scenario, for which 
asymptotic upper-bounds on the reconstruction error are provided. Both the upper and lower-bounds show that collaboration can be 
achieved through energy accumulation under certain circumstances. In order to investigate the practical performance of the proposed retransmission protocol we provide a numerical evaluation of the upper-bounds in the non-asymptotic energy regime using low-order quantization in the sensors. The latter includes a minor modification of the protocol to improve reconstruction fidelity. Numerical results show that an increase in the size of the network brings benefit in terms of performance, but that the gain in terms of energy efficiency 
diminishes quickly at finite energies due to a non-coherent combining loss.


\end{abstract}

\pagestyle{plain}
\section{Introduction \label{sec:intro}}


This paper focuses on the problem of transmitting correlated analog sources over a Gaussian multiple-access channel with a feedback link from the receiver to each encoder. The primary objective is to provide asymptotic performance measures and a realizable, simple transmission strategy for large 
one-hop sensor networks. We model systems where each sensor measures signals with a finite and small number of
source dimensions, in comparison to the number of channel dimensions.  This is motivated by applications such as remote sensing using broadband
wireless infrastructure (e.g. 4G cellular networks) where sensors take sporadic samples of a random event, feed them back to the 
network via base stations and subsequently return to an idle state to conserve power.  As a result, we do not consider coding of sequences of 
samples, but rather exploit spatial expansion and correlation between a network of sensors with independent observation noise.  
Since the applications target 4G wireless networks, it is reasonable to assume a feedback-based transmission strategy, and 
both the asymptotic results as well as the transmission strategy studied here will exploit feedback. The latter allows for simple 
and energy-efficient strategies, even if feedback is not required for optimality. 
 
The main results of this work are firstly the derivation of lower-bounds governing both the reconstruction error of a single random vector imperfectly measured 
by a network of sensors and multiple source vectors which are transmitted to a common receiver via an additive white Gaussian noise asynchronous multiple-access channel with a 
perfect causal feedback link to the encoder connected to each sensor. The bounds are expressed both for a uniform random-vector source 
with uniformly-distributed observation noise and for a Gaussian source with Gaussian observation noise. Secondly, we extend a retransmission protocol 
inspired by the classical scheme in \cite{Yamamoto79} applied to the transmission of single and bi-variate analog samples analyzed 
in \cite{EURECOM+3798} and \cite{EURECOM+3898} to the more general network with $M$ noisy observations of a common random sample. We restrict the second
analysis to uniform one-dimensional sources. The simple two-round transmission scheme combines uniform quantization and orthogonal modulation, for which 
we provide asymptotic upper-bounds on the reconstruction error as a function of the total received energy and observation noise level. Both the 
upper and lower-bounds show that a trade-off exists between the source SNR and channel SNR indicating the extent to which collaboration to be achieved 
through energy accumulation. 

Finally, we investigate the practical performance of the proposed retransmission protocol through numerical evaluation of the upper-bounds in the non-asymptotic energy regime, which corresponds to using low-order quantization in the sensors. In order to improve
the performance of the protocol, we introduce a minor modification in the feedback strategy which allows the 
error-free performance to be achieved quickly. Comparisons with a one-shot transmission not exploiting feedback are made in order to judge the benefit of the protocol in the non-asymptotic regime for a few network sizes. 

With respect to multiple-source systems, the authors in \cite{lapidoth1} and \cite{lapidoth2} derive a threshold signal-to-noise ratio (SNR) through the correlation between the sources so that below this threshold, minimum distortion is attained by uncoded transmission in a Gaussian multiple access channel with and without feedback, respectively. In these works, the authors consider transmission of a bi-variate normal source and the distortion can
be characterized by two regimes as a function of the relationship between the channel SNR and the source SNR.
In the high-correlation regime the distortion is reduced through collaboration in the received energy from the multiple-access channel and
amounts to essentially the distortion of a single-source with a factor 4 in energy efficiency. 
A similar result is provided in \cite{goblick}, whereas the latter work limits the optimality of uncoded transmission based on the SNR.
\cite{Gastpar} and \cite{oohama} can be given as further examples where collaboration has the effect of
linearly increasing the reconstruction fidelity of the source with the network size.  
In \cite{Gastpar}, however, the system parameters are chosen so that the trade-off between source and
channel collaboration is not immediately evident. 
 
The outline of the paper is as follows: in the following subsection \ref{sec:model}, we give a description of the general model to explain the problem addressed. It is followed by the derivation of the information-theoretic bounds on the reconstruction error for the two different distribution outlined above. In Section \ref{sec:mults_unif}, we provide an $M$-sensor adaptation of Yamamoto's protocol for a uniformly-distributed source with uniform observation error along with the
analysis of its asymptotic performance. In Section \ref{sec:numerical}, we present the numerical results for a slightly different protocol and the lower bounds derived in \ref{sec:est_U}. Lastly, we draw conclusions based on the results of the two analyses. 

\section{Model description} \label{sec:model}
\begin{figure}[htp]
  \centering
  \includegraphics[width=0.60\linewidth]{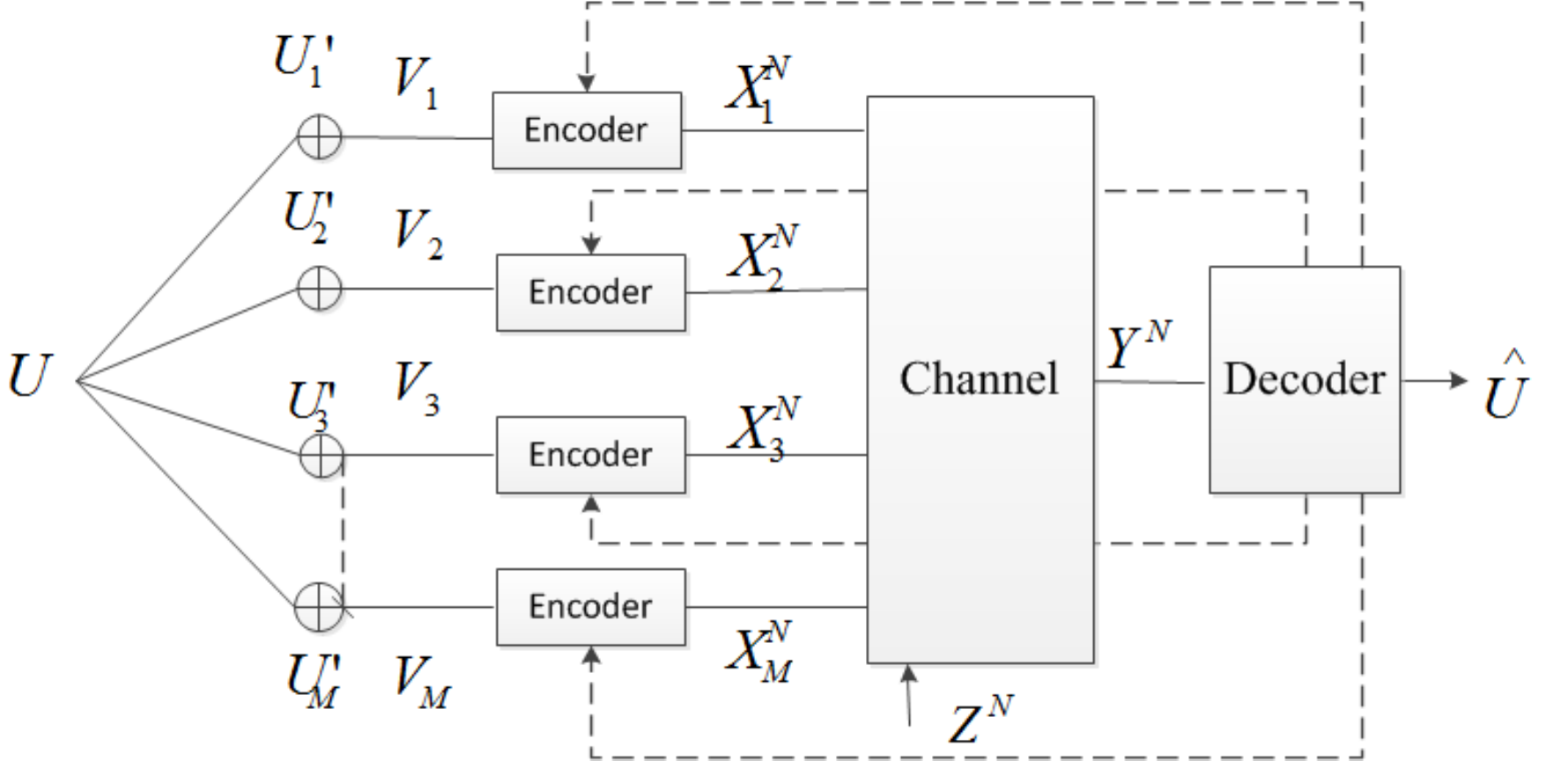}
  \caption{Pictorial representation of the described system}
	\label{fig:multiples}
\end{figure}

Let us begin with the description of the system shown in Figure \ref{fig:multiples}.
In order to extend the system studied in \cite{EURECOM+4081} as the source model II for dual-sources, we will describe an adaptation to a general model which includes the first source as the mutual element of the whole system and combined with $M$ other sources through a correlational relationship. 
The construction of the sources is given by the following linear expression.
\begin{equation}\label{eq:multso}
\mathbf{V}_{j}=\rho \mathbf{U}+\sqrt{1-\rho^2} \mathbf{U}_{j}'
\end{equation}
Here we denote the $M$ auxiliary random vectors representing the observation noise in each sensor by $\mathbf{U}_{j}'$ and the observation of the  
mutual source $\mathbf{U}$ by $\mathbf{V}_{j}$, both of dimension $K$, for $j=1,2,...,M$. Each realization of $\mathbf{V}_{j}$ is mapped into $\mathbf{X} \triangleq (X_1,\ldots,X_N)$ which is then sent across the channel 
corrupted by a white complex circularly symmetric Gaussian noise sequence $\mathbf{Z}$, and is received as the output signal $\mathbf{Y}$. The receiver constructs an estimate $\widehat{\mathbf{U}}$ of $\mathbf{U}$ given $\mathbf{Y}$. The transmitted sequence $\mathbf{X}$ is encoded as $X_i=f_{i,j}(\mathbf{V}_{j},Y_1,\cdots,Y_{i-1})$, where the function $f_{i,j}$ is an arbitrary mapping for the $j^{\mathrm{th}}$ sensor in dimension $i$ and depends on perfect knowledge of past observations.  The latter models an ideal causal feedback path from the receiver. The dimension of the channel input is denoted by $N$ and can be assumed to be large, whereas $K$ is assumed to be finite and small. 
 
We consider two cases for the distribution of $\mathbf{U}$. In the first case, both the $\mathbf{U}_{j}'$ and $\mathbf{U}$ are uniformly distributed with zero mean and unit variance, i.e. defined in the range $(-\sqrt{3},\sqrt{3})$. Depending on the level of correlation, $\mathbf{V}_{j}$ defined by (\ref{eq:multso}) has a contaminated uniform distribution. We will consider the case where $\mathbf{V}_{j}$, $\mathbf{U}$ and $\mathbf{U}_{j}'$ are standard normally distributed which is equivalent to having the parameters $\mathcal{N}(0,1)$. The output signal and the power constraints are given in the following by
\begin{equation}
Y_{i}=\sum_{j=1}^{M} X_{i,j}e^{i\phi_{i,j}}+Z_{i},\;\;
\frac{1}{K}\sum_{i=1}^{N} E[|X_{i,j}|^2]\leq \mathcal{E}_{j}
\end{equation}
for $j=1,2,...,M$ and $i=1,...,N$, respectively.
The criteria for source-channel code design is chosen as the squared-error distortion measure, which is $d(u_{i},\hat{u}_{i})=(u_{i}-\hat{u}_{i})^2$ for $i=1,2,\cdots K$, and the average distortion is defined as
\begin{equation}
D = \frac{1}{K}\mathrm{E}\left[\sum_{k=1}^{K}d(u_i,\hat{u})\right]. 
\end{equation} $\mathbf{\phi}_{j}=\{\phi_{j,i};i=1,...,N\}$
denotes the random phase sequences which are assumed to be i.i.d. uniform over $[0,2\pi)$ and 
unknown to the transmitter and receiver.  The latter models an asynchronous network and the fact that a coherent reception model is unrealistic for sporadic
information transfer. These assumptions are implicitly relaxed in the lower bounds on the distortion discussed in the following section 
but are applied in the coding strategy considered in the upcoming section.
\section{Estimation of $U$ \label{sec:est_U}}
In order to obtain a bound on the fidelity of estimating the random vector $\mathbf{U}$, we obtain upper and lower bounds on a cut-set mutual information
functional $I(\mathbf{U};\mathbf{Y}|\{\mathbf{V}_{j}\}_{S})$ based on a subset $S\subseteq{1,2,\cdots,M}$ and its complement $S^{c}$. 
$\{\mathbf{V}_{j}\}_{S}$ denotes the subset of $\mathbf{V}_{j}$'s for $j\in S$. The derivations of the two bounds are given in the Appendix \ref{subsec:appI} for both
uniform and normal distributions. The bounds are summarized as
\begin{equation}
I(\mathbf{U};\mathbf{Y}|\{\mathbf{V}_{j}\}_{S})\geq-h(\{\mathbf{V}_{j}\}_{S})+h(\{\mathbf{V}_{j}\}_{S}|\mathbf{U})
+h(\mathbf{U})-h(\mathbf{U}-\mathbf{\hat{U}}).
\end{equation}
\begin{equation}
I(\mathbf{U};\mathbf{Y}|\{\mathbf{V}_{j}\}_{S})\leq N\log \left(1+\frac{K\sum_{j\in S^{c}}\mathcal{E}_{j}}{N N_0}\right).
\end{equation}

Combining the two bounds given above allows us to express the distortion level for estimating the mutual random vector $\mathbf{U}$ as
\begin{equation}  \label{ineq:dist_general}
D\geq \max_{|S|}C_D \frac{1-\rho^2}{1+(|S|-1)\rho^2}\left(1+\frac{K\sum_{j\in S^{c}}\mathcal{E}_{j}}{NN_0}\right)^{-\frac{2N}{K}}
\end{equation}
where $C_D$ is a constant which varies based on the distribution type and defined as
\begin{equation*} \label{constant}
C_{D}=
\begin{cases}
(\frac{6}{\pi e})^{|S|+1}, & \text{for}\;\;U\sim\mathcal{U}(-\sqrt{3},\sqrt{3})\\
1, & \text{for}\;\;U\sim\mathcal{N}(0,1).\\
\end{cases}
\end{equation*}
The general bound given above by (\ref{ineq:dist_general}) includes two parameters; the correlation coefficient $\rho$ 
and the energy term and is valid for all $0\leq |S|\leq M$. 
In the source-channel coding scheme proposed in the following section which targets broadband networks and small
amounts of analog information, we are mostly interested in the case where 
$N\gg K$, or where the channel bandwidth is significantly higher than the source bandwidth.   
For $N\to \infty$ and $\mathcal{E}_j=\mathcal{E}\;\;\forall j$, (\ref{ineq:dist_general}) becomes
\begin{equation} \label{ineq:multd}
D\geq\max_{|S|}C_D \frac{1-\rho^2}{1+(|S|-1)\rho^2} \exp{\left(-\frac{2(M-|S|)\mathcal{E}}{N_0}\right)}.
\end{equation}
which can easily be simplified to
\begin{equation}\label{ineq:multd2}
D\geq\begin{cases}
C_D \left(\frac{1-\rho^2}{1+(M-1) \rho^2}\right), & \frac{1+(M-2)\rho^2}{1+(M-1)\rho^2} \geq e^{-\frac{2\mathcal{E}}{N_0}} \\
...                                                                     \\                                                                     \\
C_D \left(\frac{1-\rho^2}{1+(i-1) \rho^2}\right)e^{-\frac{2(M-i)\mathcal{E}}{N_0}}, & \frac{1+(M-i-1)\rho^2}{1+(M-i)\rho^2}  \geq e^{-\frac{2\mathcal{E}}{N_0}} \\
... \\
C_D\exp{\left(-\frac{2M\mathcal{E}}{N_0}\right)}, & 1-\rho^2 \le e^{-\frac{2\mathcal{E}}{N_0}}.
\end{cases}
\end{equation}
The above result brings to light the effect of collaboration between the sensors which is achieved either through
the spatial expansion in the channel or in the source. To see this, we note that the condition for the $i^{th}$ source
$\frac{1+(M-i-1)\rho^2}{1+(M-i)\rho^2}  \geq e^{-\frac{2\mathcal{E}}{N_0}}$ is equivalent to saying that the 
distortion in each sensor node induced by the observation process is more significant than the lowest distortion offered by
the channel when estimating $\mathbf{V}_j$ (which is $D_c \geq e^{-\frac{2\mathcal{E}}{N_0}}$) in the absence of the signals from the other sensors.  
Note that this is the classical point-to-point optimal distortion derived in \cite{goblick}. 

Let us consider the vector channel model $\mathbf{Y}=\sqrt{snr}\mathbf{H}\mathbf{X}+\mathbf{N}$ analyzed in \cite{Verdu}, where $\mathbf{X}$ and $\mathbf{Y}$ are the input and output signals, $\mathbf{H}$ is a deterministic matrix and $\mathbf{N}$ represent the channel noise in the described model. The source component of our system (i.e. up to the input of the channel encoder) can be simply considered as a special-case of the estimation problem treated in \cite[eq.19]{Verdu} through the following definition for the Gaussian construction,
\begin{equation}\label{model_alt}
\mathbf{Y}'=\frac{\rho}{\sqrt{1-\rho^2}}\mathbf{H}\mathbf{X}+\mathbf{N}.
\end{equation} Here in (\ref{model_alt}), our mutual source $\mathbf{U}$ from the model (\ref{eq:multso}) is replaced by the vector $\mathbf{X}$, the auxiliary random vector $\mathbf{U}'_j$ in other words the observation noise is represented by the channel noise $\mathbf{N}$, where the output signal $\mathbf{Y}'=\mathbf{Y}/\sqrt{1-\rho^2}$ corresponds to the vector of $\mathbf{V}_j$'s in our model. Attaining the corresponding vectors, we obtain the mean square error in estimating mutual source $\mathbf{U}$ given $\mathbf{V}_j$ which is no different than estimating $\mathbf{H}\mathbf{X}$ in the original work  as
\begin{equation}
\mathrm{E}\left\{||\mathbf{U}-\hat{\mathbf{U}}||^2|\mathbf{V}_j\right\}=\frac{1-\rho^2}{1+(M-1)\rho^2}
\end{equation}
since $\mathbf{H}$ is a matrix with all elements are 1. 

A comparable trade-off regarding the collaboration effect due to the source or channel can be seen in \cite{lapidoth1,lapidoth2} for the case
$K=N$.  
As mentioned in the introduction, another example is the Gaussian sensor network application \cite[sections VI and VII]{Gastpar} (again for $K=N$) 
or the CEO problem studied in \cite{oohama}, where estimation fidelity decays
linearly with the size of the network in a manner similar to (\ref{ineq:dist_general}).

\section{Estimation of the set of $V_j$'s \label{sec:est_V}}
\subsection{Bound on some subset of $V_j$'s \label{sec:bound_V}}

As mentioned above, another way of approaching to the multiple-source problem is the estimation of $\mathbf{V}_{j}$'s instead of the mutual element $\mathbf{U}$. The mutual information that is used to obtain the lower bound on  the reconstruction error is $I(\mathbf{V}_{j};\mathbf{Y}|\{\mathbf{V}_{l}\}_{S})$ where $\{\mathbf{V}_{l}\}_{S}$ denotes the set of $V_{j}$'s (any subset to be chosen) which excludes $j$, i.e. $\{\mathbf{V}_{l}\in S, S\subset \{1,...,M\}-j \}$. First expansion of $I(\mathbf{V}_{j};\mathbf{Y}|\{\mathbf{V}_{l}\}_{S})$ based on the output signal proceeds in the same way for both distribution types as follows
\begin{equation} \label{mults_E'}
I(\mathbf{V}_{j};\mathbf{Y}|\{\mathbf{V}_{l}\}_{S})\leq N\log \left(1+\frac{K\sum_{j\in S^c}\mathcal{E}_{j}}{N N_0}\right).
\end{equation} 
We obtain the following second expansion based on the source entropies
\begin{equation}
I(\mathbf{V}_{j};\mathbf{Y}|\{\mathbf{V}_{l}\}_{S})\geq -h(\{\mathbf{V}_{l}\}_{S})+h(\{\mathbf{V}_{l}\}_{S}|\mathbf{V}_{j})+h(\mathbf{V}_{j})-h(\mathbf{V}_{j}-\mathbf{\hat{V}}_{j}) \label{mults_F'}
\end{equation}
The expansion on the output signal given by (\ref{mults_E'}) which at the end includes the energy term is independent of the source distribution type. But clearly, (\ref{mults_F'}) differs as given in the Appendix \ref{subsec:appII}.
The two expressions of the same mutual information $I(\mathbf{V}_{j};\mathbf{Y}|\{\mathbf{V}_{l}\}_{S})$ given by (\ref{mults_E'}) and (\ref{mults_F'}) are used to obtain the following lower-bound on the distortion.
\begin{equation} \label{mults_G3}
D \geq C_D \left(1+\frac{K\sum_{j\in S^c}\mathcal{E}_{j}}{NN_0}\right)^{-\frac{2N}{K}} 
\end{equation}
Asymptotically in $N$, we obtain
\begin{equation} \label{ineq:multdu1}
D \geq C_D \exp{\left(-\frac{2(M-|S|)\mathcal{E}_{j}}{N_0}\right)}
\end{equation}
where $C_{D}$ is a constant differs based on the source distribution and given as
\begin{equation*} \label{constant_V}
C_{D}=
\begin{cases}
\frac{(1-\rho^2)}{1+(|S|-1)\rho^2}(2-\rho^2+|S|), & \text{for}\;\; \text{Gaussian}\\
\frac{12^3 \rho^2(1-\rho^2)^{2-|S|}}{(2\pi e)^{|S|+2}(1+(|S|-1)\rho^2)}+\left(\frac{6}{\pi e}\right)^{|S|+1}\frac{1-\rho^2}{1+(|S|-1)\rho^2}, & \text{for}\;\; \text{Uniform.}\\
\end{cases}
\end{equation*}


\subsection{Bound on product distortion $\prod_{j=1}^{M} D_j $ \label{sec:product_V}}

In addition to the lower bound (\ref{ineq:multdu1}) on the reconstruction error of estimating $V_j$'s introduced in the previous subsection, the product distortion $D_{1}D_{2}...D_{M}$ can be also bounded using two different expansions of $I({\mathbf{V}_{j}};\mathbf{Y})$, where ${\mathbf{V}_{j}}$ is the whole set with $j=1,2,...,M$. The resulting lower bound on $\prod_{j=1}^{M}D_j$ is obtained as
\begin{equation} \label{ineq:prod_M}
\prod_{j=1}^{M} D_j \geq C_p \left(1+\frac{KM \mathcal{E}}{NN_0}\right)^{-2N/K}
\end{equation} 
with
\begin{equation*} \label{constant_prod}
C_{p}=
\begin{cases}
(1-\rho^2)^M (1+\frac{M \rho^2}{1-\rho^2}), & \text{for}\;\; \text{Gaussian}\\
\left(\frac{(12 \rho^2)^{1/M}+12(1-\rho^2)}{2\pi e}\right)^M, & \text{for}\;\; \text{Uniform.}\\
\end{cases}
\end{equation*}
where $\mathcal{E}_j=\mathcal{E} \;\;\forall j$. Let $N \to \infty$, above expression becomes
\begin{equation} \label{ineq:prod_M_asym}
\prod_{j=1}^{M}D_j\geq C_p \exp{\left(-\frac{2M\mathcal{E}}{N_0}\right)}.
\end{equation} Lower bound (\ref{ineq:prod_M_asym}) can be simplified to $D \geq (C_p)^{1/M} \exp{\left(-\frac{2 \mathcal{E}}{N_0}\right)}$ by taking the $M^{th}$ root for equal distortions per source, i.e. $D_j=D\;\;\forall j$,  which is the Goblick bound \cite{goblick} achieved for a point to point channel.
The derivations are given in detail in Appendix \ref{subsec:appIV}.
\section{Achievable Scheme for a network with Uniform sources \label{sec:mults_unif}}

The two-way protocol introduced in \cite{EURECOM+3798} for a single source and its extension to dual-source studied in detail in \cite{EURECOM+3898} is generalized to large networks where the same approach is applied to a scheme with $M$ sources for $M\geq 2$. 
As depicted in Figure \ref{fig:multiples}, there is one mutual source which is represented by $U$, and $M$ other auxiliary random variables which are combined in pairs linearly through (\ref{eq:multso}) each of which includes $U$ through a correlational relationship and are distributed uniformly included within the range $(-\sqrt{3},\sqrt{3})$.
\begin{figure}[htp]
  \centering
  \includegraphics[width=0.40\linewidth]{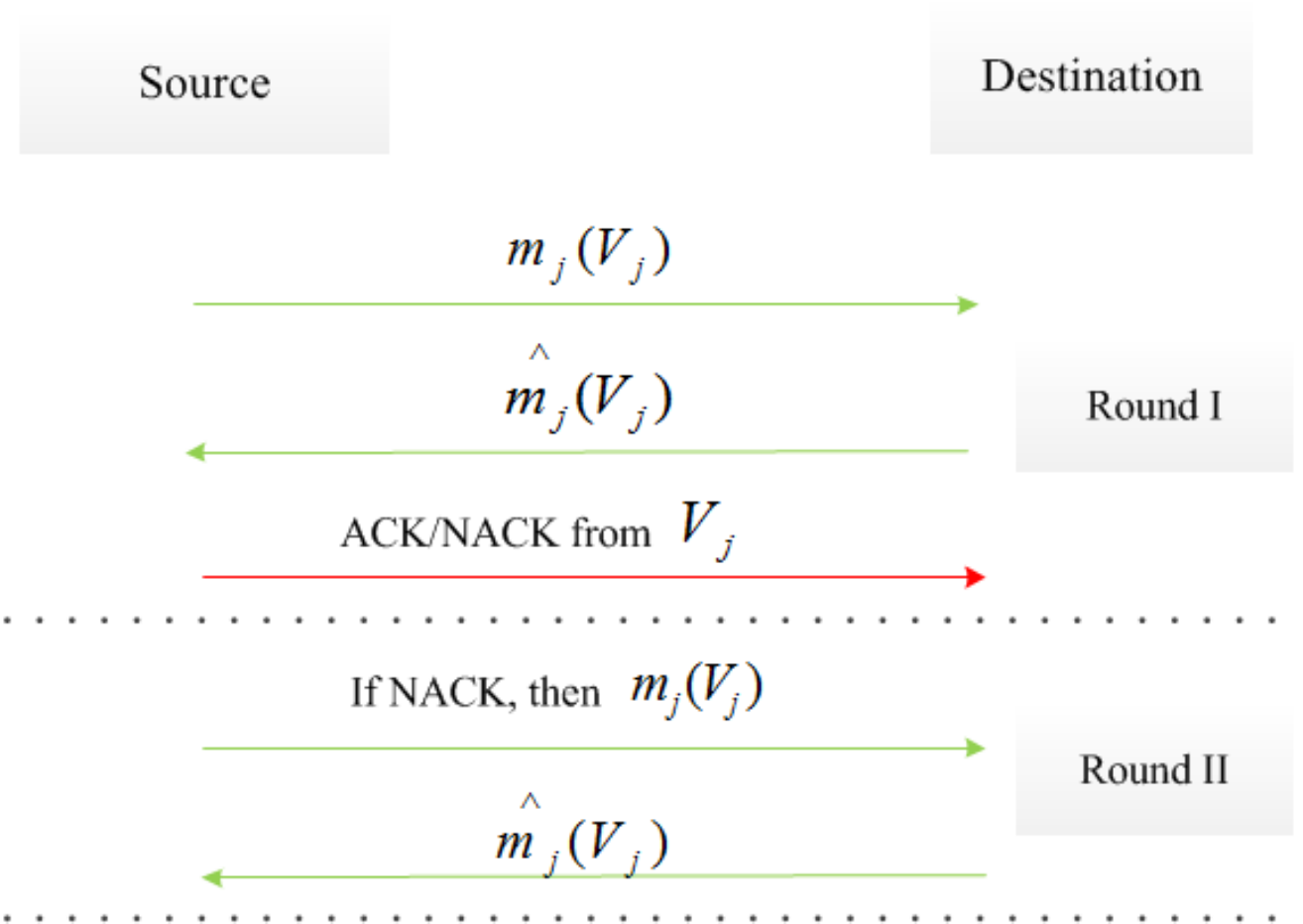}
  \caption{Two-round protocol}
	\label{fig:protocol_M}
\end{figure}

\begin{figure}[htp]
  \centering
  \includegraphics[width=0.70\linewidth]{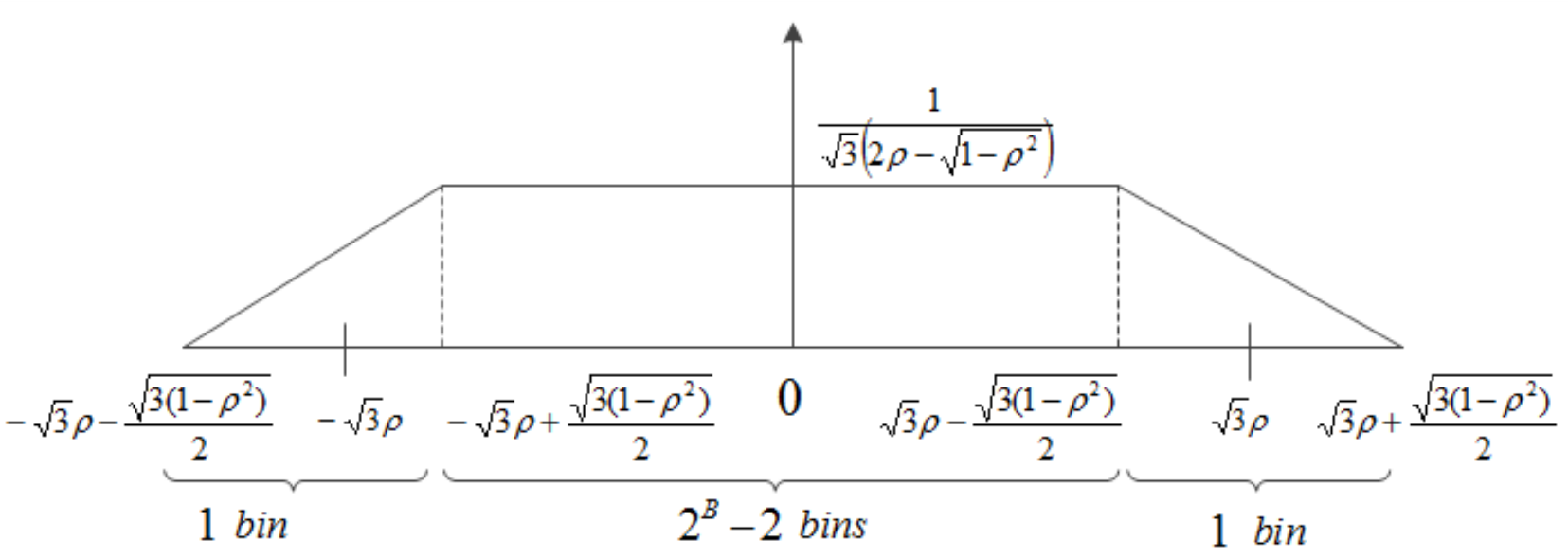}
  \caption{Pictorial representation of quantization process for the defined distribution with the allocation of the quantization bins}
	\label{fig:quantizer}
\end{figure}

\begin{figure}[htp]
  \centering
  \includegraphics[width=0.35\linewidth]{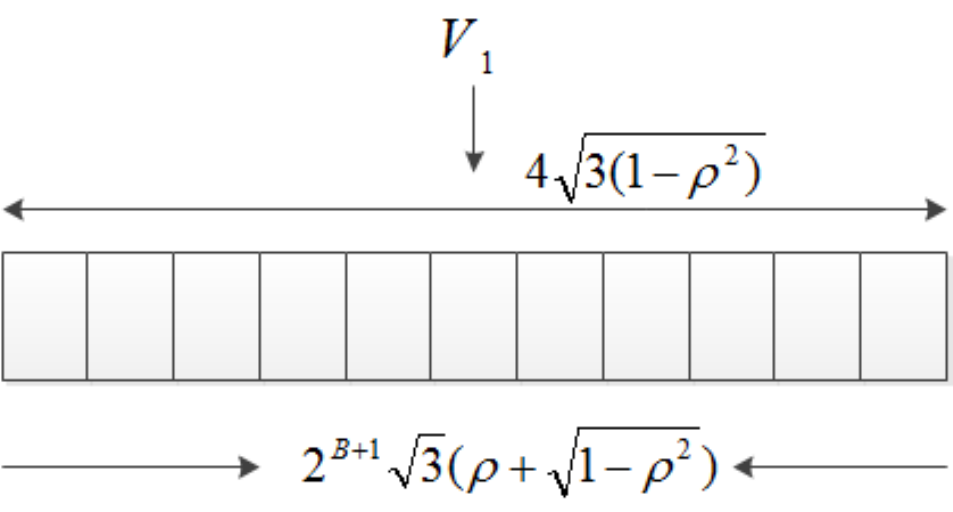}
  \caption{Pictorial representation of detection }
	\label{fig:detector}
\end{figure}

The protocol consists of two phases which composes one round and proceeds as depicted in Figure \ref{fig:protocol_M}. First phase is called the data phase, in which the first transmission occurs and in return feedback of the messages are received from the decoder by each encoder. The quantization process is depicted in Figure \ref{fig:quantizer}. We assume that the source sample of the $j^{th}$ source which is uniformly quantized is subsequently encoded into $2^{B_{j}}$ messages with dimension $N$ where $B_{j}$'s are equal to the same value $B$. Each tail of the distribution is considered as one quantization bin and the interior part, which consists of $2^B-2$ bins, is uniformly quantized. Note that, for a full correlation between the sources, i.e. $\rho=1$,  the 'contamination' in the source distribution vanishes and the shape given by Figure \ref{fig:quantizer} becomes a rectangular.
We fix the total energy that is used by protocol and for the sake of simplicity the energy used in one round is allocated equally among $V_{j}$'s for $j=1,2,...,M$, e.g. for the data phase of the first round the aggregate energy is denoted by $\mathcal{E}_{\mathrm{D},1}$ where $\mathcal{E}_{\mathrm{D},1}=\sum_{j=1}^{M}\mathcal{E}_{\mathrm{D},1,j}$. 
The quantized source sample of the $j^{th}$ source is encoded into $2^{B_{j}}$ messages with dimension $N$ where $B_{j}$'s are equal to the same value $B$. The chosen method is $2^{B}$-ary orthogonal modulation with non-coherent reception. In the data phase, the $j^{th}$source sends its message ${m}_j=Q(V_{j})$ to the receiver with energy $\mathcal{E}_{\mathrm{D},1,j}$. The aggregated source messages are denoted by $\mathbf{m}$ which is a vector of the messages $(m_1, m_2,..., m_M)$ with dimension $M$. Note that, all messages from different sources are mutually orthogonal. The receiver decodes $\hat{m}_{j}$ and feeds it back.  
The output signal based on the $N$ dimensional observation of the $j^{th}$ source is given as  
\begin{equation}
\mathbf{Y}_{d_{j}} = \sqrt{\mathcal{E}_{\mathrm{D},1,j}}\mathrm{e}^{j\Phi_j}\mathbf{S}_{m_{j}} + \mathbf{Z}_{d_{j}}.
\end{equation}
We assume the random phases $\Phi_{j}$ to be distributed uniformly on $[0,2\pi)$, the channel noise $\mathbf{Z}_{d_{j}}$ to have zero mean and equal autocorrelation $N_{0}\mathbf{I}_{N\times N}$ for $j=1,2,...,M$ and  $\mathbf{S}_{m_{j}}$ are the $N$-dimensional messages, with $m=1,2,\cdots, 2^B$. At the receiver end, we consider the following exhaustive search. To decode the first $m_{1}$, there are $2^B$ possibilities whereas $m_{j>1}$ is constrained to $2^{B}(\frac{2\sqrt{1-\rho^2}}{\rho+\sqrt{1-\rho^2}})$ since it cannot fall outside of the interval $V_{1}+\sqrt{1-\rho^2}(U_{j}'-U_{1}')$ which is depicted in Figure \ref{fig:detector}. 
The detection rule is given using \cite[Chapter 12]{ProakisBook} considering the following $M$ possible decision variables assuming $(m_1, m_2,..., m_M)$ is transmitted where 
\begin{align} \label{dec_var_M}
\mathbf{U}_{\mathbf{m}'}&=\sum_{j=1}^{M}|<\mathbf{Y}_j,\mathbf{S}_{m_{j}}>|^2 \nonumber \\
& = \sum_{j: m_j=m'_j} |\sqrt{\mathcal{E}_{D,1,j}}+N_j|^2+ \sum_{j: m_j\neq m'_j} |N_j|^2
\end{align}

According to (\ref{dec_var_M}), the receiver chooses $\hat{\mathbf{m}}=\mathrm{argmax}_{\hat{\mathbf{m'}}}\;\mathbf{U}_{\mathbf{m'}}$.
After the estimation and feedback of $\hat{m}_j$ to each encoder, data phase of the first round ends and the encoders enter the control phase to inform the receiver about the correctness of its decision by sending ACK/NACK regarding its own message to the decoder. 
During the control phase the receiver observes
$\mathbf{Y}_{c}$ with $\mathbf{Y}_{c_{j}} = \sqrt{\mathcal{E}_{\mathrm{C},1,j}}\mathbf{A}_{j}\mathrm{e}^{j\Phi_j}\mathbf{S}_{c_{j}} + \mathbf{Z}_{c_{j}}$ for $j^{th}$ source where $\mathbf{A}_{j}$ takes the value 0 for a signal ACK and 1 for a NACK and $\mathcal{E}_{\mathrm{C},1,j}$ here denotes the energy of the control phase in the first round on one source. So the encoders inform the receiver whether or not its decision 
was correct via a signal $\sqrt{\mathcal{E}_{\mathrm{C},1,j}}\mathbf{S}_{c_{j}}$ of energy $\sqrt{\mathcal{E}_{\mathrm{C},1,j}}$ if the
decision is incorrect and $\mathbf{0}$ if the decision was correct.
The detector defined in \cite{EURECOM+3898} is adapted to the current scheme and given for the $j^{th}$ source as 
\begin{equation}
e_{j}=I\left(|y_{c,j}|^{2}>\lambda \mathcal{E}_{\mathrm{C},1,j}\right)
\end{equation} 
with $y_{c,j}=\mathbf{Y}_{c,j}^H\mathbf{S}_{c,j}$ where $I(\cdot)$ is the indicator function and $\lambda$ is a threshold to be optimized that is included within the interval $[0,1)$.
$\Pr(E_{e\rightarrow c,1}|k\;\;in\;\;error)$ denotes the total probability of uncorrectable error given that $k$ sources are in error in the first round where the probability for a single source is bounded by the recent bound introduced in \cite[eq. 12]{Kam06}
\begin{align}
\Pr(E_{e \rightarrow c,1,j})&=\Pr(|\sqrt{\mathcal{E}_{\mathrm{C},1,j}}+z_{c,j}|^2 \leq \lambda\mathcal{E}_{\mathrm{C},1,j}) \nonumber \\
&=1-Q_{1}\left(\sqrt{\frac{\mathcal{E}_{\mathrm{C},1,j}}{N_{0}/2}},\sqrt{\frac{\lambda\mathcal{E}_{\mathrm{C},1,j}}{N_{0}/2}}\right)\nonumber \\
&\overset{(a)}{\leq}1/2 \exp\left(-\frac{(\sqrt{\lambda}-1)^2 \mathcal{E}_{\mathrm{C},1,j}}{N_{0}}\right). \label{probec_M}
\end{align} Using the bound given above, the probability yields for $k$ sources 
\begin{equation}
\Pr(E_{e\rightarrow c,1}|k\;\;in\;\;error)\leq \left(\frac{1}{2}\right)^k  \exp \left(-\frac{k(\sqrt{\lambda}-1)^2 \mathcal{E}_{\mathrm{C},1}}{MN_{0}}\right)
\end{equation} where $\mathcal{E}_{\mathrm{C},1,j}=\mathcal{E}_{\mathrm{C},1}/M \;\;\forall j$.
In the case of at least one NACK out of $M$ control signals is received, the protocol goes on one more round for retransmission, otherwise it is terminated. And the second data phase starts after the sources are instructed by the destination in order to do the retransmission. The scheme could be generalized to more than two rounds.
The union bound on $P_e(\mathbf{m})$, the probability of error at the end of the first round, is given by
\begin{align} \label{union_M}
P_e(\mathbf{m})&\leq \sum_{ \mathbf{m}\neq \mathbf{m}'} \Pr(\mathbf{U}_{\mathbf{m}}<\mathbf{U}_{\mathbf{m'}}|\mathbf{m})  \nonumber \\
&= 2^{B}\left\lceil 2^{B+1}\frac{\sqrt{1-\rho^2}}{\rho+\sqrt{1-\rho^2}}\right\rceil ^{M-1} P_{2}(M)+\sum_{k=1}^{M-1} \dbinom{M}{k} \left\lceil 2^{B+1}\frac{\sqrt{1-\rho^2}}{\rho+\sqrt{1-\rho^2}}\right\rceil ^{k} P_{2}(k).
\end{align}

The decision variables (\ref{dec_var_M}) are used to bound the conditional probability (\ref{union_M}). $P_{2}(k)$ is defined in \cite{ProakisBook} through the following equality
\begin{equation}
P_{2}(k)= \frac{1}{2^{2k-1}} e^{-\gamma} C_{k} \label{eq:p2}
\end{equation}
with $C_k$ defined as
\begin{equation} \label{C_factor}
C_{k}=\sum_{n=0}^{k-1} \left(\frac{1}{n!}\sum_{l=0}^{k-1-n}\dbinom{2k-1}{l}\right) \gamma^{n}
\end{equation}
where $\gamma$ denotes the SNR. Second round decision variables are represented by $U_{\mathbf{m'}}^{(2)}$ and given by
\begin{equation} \label{dec_var_M2}
\mathbf{U}_{\mathbf{m'}}^{(2)}=\mathbf{U}_{\mathbf{m'}}+\sum_{j=1}^{M}|<\mathbf{Y}_{j}^{(2)},\mathbf{S}_{m_{j}}>|^2
\end{equation} 
This is analogous to soft or chase-combining in HARQ mechanisms.
As in the first round, the receiver chooses $\hat{\mathbf{m}}=\mathrm{argmax}_{\hat{\mathbf{m'}}}\;\mathbf{U}_{\mathbf{m'}}^{(2)}$ over all possible sequences, thereby disregarding the messages which were hypothesized to be correct after the control phase. 
Using the chosen estimator $\hat{u}=\frac{1}{M}\sum_{i=1}^{M}\hat{v}_j/ \rho$, the protocol terminates with the following distortion at the end of the second round.
\begin{equation} \label{gendist_M}
D(\mathcal{E},N_{0},2,\lambda) \leq D_q+\sum_{k=1}^{M-1} D_{e,k}P_{e,k}+D_{e,M} P_{e,M} 
\end{equation} The derivation of the bound (\ref{gendist_M}) can be found in Appendix \ref{subsec:appIII}.
Here, $k$ corresponds to the number of the sources in error with the values $k=1,...,M-1$.
The error probability corresponding to the case where at least one out of $M$ sources being correct is represented in the above expression by $P_{e,k}$ which is the case of any of $k$ sources being in error including the uncorrectable error after the first round, or $k$ being in error at the end of the second round. 
$P_{e,M}$ represents all of the $M$ sources being in error after the first or second round. 
Note that, $P_{e,k}$ which consists of $P_{2}(2k)$ and $P_{2}(k)$ is given by 
\begin{align} \label{eq:pek}
P_{e,k}&\leq \sum_{\substack{\mathbf{m}' \neq \mathbf{m} \\ d_{H}(\mathbf{m}',\mathbf{m})=k}}\left(\Pr(\mathbf{U}_{\mathbf{m}}<\mathbf{U}_{\mathbf{m'}}|\mathbf{m})\Pr(E_{e \rightarrow c,1}|k\;\;in\;\;error)+\Pr(\mathbf{U}_{\mathbf{m}}^{(2)}<\mathbf{U}^{(2)}_{\mathbf{m'}}|\mathbf{m})\right)\nonumber \\
&=\dbinom{M}{k}\left(\left\lceil 2^{B+1}\frac{\sqrt{1-\rho^2}}{\rho+\sqrt{1-\rho^2}}\right\rceil ^{k} \Pr(E_{e \rightarrow c,1}|k\;\;in\;\;error) P_{2}(k) +\left\lceil 2^{B+1}\frac{\sqrt{1-\rho^2}}{\rho+\sqrt{1-\rho^2}}\right\rceil ^{k} P_{2}(2k)\right)
\end{align} where $d_{H}(\mathbf{x},\mathbf{y})$ denotes the Hamming distance between two vectors $\mathbf{x}$ and $\mathbf{y}$.
In the same way, $P_{2}(M)$ and $P_{2}(2M)$ shape together $P_{e,M}$ as given in the following form. 
\begin{align} \label{eq:pem}
P_{e,M}&\leq  \sum_{\substack{\mathbf{m}' \neq \mathbf{m} \\ d_{H}(\mathbf{m}',\mathbf{m})=M}}\left(\Pr(\mathbf{U}_{\mathbf{m}}<\mathbf{U}_{\mathbf{m'}}|\mathbf{m})\Pr(E_{e \rightarrow c,1}|M\;\;in\;\;error)+\Pr(\mathbf{U}_{\mathbf{m}}^{(2)}<\mathbf{U}^{(2)}_{\mathbf{m'}}|\mathbf{m})\right)\nonumber \\
&= 2^{B}\left\lceil 2^{B+1}\frac{\sqrt{1-\rho^2}}{\rho+\sqrt{1-\rho^2}}\right\rceil ^{M-1} \Pr(E_{e \rightarrow c,1}|M\;\;in\;\;error) P_{2}(M) +2^{B}\left\lceil 2^{B+1}\frac{\sqrt{1-\rho^2}}{\rho+\sqrt{1-\rho^2}}\right\rceil ^{M-1} P_{2}(2M)
\end{align} 
Accordingly, the corresponding distortion terms are denoted by $D_{e,k}$ and $D_{e,M}$, respectively.
When $\mathbf{m}$ is decoded correctly, the reconstruction error $D_q$ is caused solely by the quantization process and source observation error. Let us denote the estimation error by $e$, so that its variance $E [|u-\hat{u}|^2 l\;in\;error]$ for $l=0$ yields the quantization distortion with the following expansion,
\begin{align}\label{D_q}
&D_q =E\left[\frac{1}{\rho M}\sum_{j=1}^{M}\left(\sqrt{1-\rho^2}u'_j+e_{q,j}\right)\right]^2\nonumber \\ 
&=\frac{1-\rho^2}{\rho^2 M}+\frac{1}{\rho^2 M}Var(e_q)+\frac{2\sqrt{1-\rho^2}}{\rho^2 M^2}\sum_{}^{}E[u'_j e_{q,j}]\nonumber \\
&\leq \frac{1-\rho^2}{\rho^2 M}+\frac{\sqrt{3}}{\rho^2 M}\left(2^{-2 B+1}+\sqrt{3}(1-\rho^2)/2\right)+\frac{2^{-B+2}3\sqrt{1-\rho^2}}{M\rho^2}
\end{align} where $e_{q}=\frac{1}{\rho M}\sum_{j=1}^{M} e_{q,j}$ denotes the estimation error. 
The squared distortion when $k$ out of $M$ sources are decoded in error at the end of the second round is calculated through $D_{e,k}=E\left[|u-\hat{u}|^2 k\;in\;error\right]$ for $k=1,2,...,M-1$ by using the chosen estimator expanded as
\begin{align}\label{est_k}
D_{e,k}&= E\left[\frac{1}{\rho^2 M^2}\left(\sum_{\substack{j\;s.t. \\ \hat{v}_{j}\neq v_{j}}}(\rho u-\hat{v}_{j})+\sum_{\substack{j\;s.t. \\ \hat{v}_{j}=v_{j}}}(\rho u-\hat{v}_{j})\right)^2\right]\nonumber \\
&\leq \frac{(M+8k)(1-\rho^2)+2^{-B+2}((M+2k)\sqrt{1-\rho^2}+M 2^{-B})}{M^2 \rho^2/3}
\end{align}
and bounded considering the furthest distances between $u$ and its estimate for the cases when $\hat{v}_j$ is correctly and incorrectly decoded.  Lastly, $D_{e,M}=E\left[|u-\hat{u}|^2 M\;in\;error\right]$ is expanded by
\begin{align} \label{est_M}
D_{e,M}&=E \left(\frac{1}{\rho M}\sum_{j=1}^{M} \rho u-\hat{v}_{j}\right)^2\nonumber \\
&\leq 1+\frac{12}{M}+\frac{12\sqrt{1-\rho^2}}{\rho M}+\frac{3(1-\rho^2)}{\rho^2 M}
\end{align} and bounded again through the peak distortion. 
Note that, $D_q$ and $D_{e,k}$ ($1<k<M$) are in the exponential order of $2^{-2B}$ while $D_{e,M}$ is upper bounded by an order of 1.
Using the above defined error probabilities the following bound on the distortion is achieved
 \begin{align}
&D\leq D_q
+ D_{e,M} \left(K_{1}\frac{\sqrt{1-\rho^2}}{\rho+\sqrt{1-\rho^2}}e^{(B+1)\ln2}+K_2 \epsilon(\rho)\right)^{M-1}e^{(B-2M+2)\ln2-\frac{\mathcal{E}_{\mathrm{D},1}+2\mathcal{E}_{\mathrm{C},1}(\sqrt{\lambda}-1)^2}{2N_{0}}} \nonumber \\
&+ D_{e,M} \left(K_{3}\frac{\sqrt{1-\rho^2}}{\rho+\sqrt{1-\rho^2}}e^{(B+1)\ln2}+K_4 \epsilon(\rho)\right)^{M-1}e^{(B-4M+2)\ln2-\frac{\mathcal{E}_{\mathrm{D},1}+\mathcal{E}_{\mathrm{D},2}}{2N_{0}}}\nonumber \\
&+\sum_{k=1}^{M-1} D_{e,k} \dbinom{M}{k}  \left(K_{5}(k)\frac{\sqrt{1-\rho^2}}{\rho+\sqrt{1-\rho^2}}e^{(B+1)\ln2}+K_6(k)\epsilon(\rho)\right)^{k}e^{-\frac{k(\mathcal{E}_{\mathrm{D},1}+2\mathcal{E}_{\mathrm{C},1}(\sqrt{\lambda}-1)^2)}{2MN_0}} \nonumber \\
&+\sum_{k=1}^{M-1} D_{e,k} \dbinom{M}{k} \left(K_{7}(k)\frac{\sqrt{1-\rho^2}}{\rho+\sqrt{1-\rho^2}}e^{(B+1)\ln2}+K_8(k)\epsilon(\rho)\right)^{k}e^{-\frac{k(\mathcal{E}_{\mathrm{D},1}+\mathcal{E}_{\mathrm{D},2})}{2MN_0}} \label{dist_M}
\end{align} where $K_{1},K_{2}$ are $O((\mathcal{E}_{\mathrm{D},1})^{M-1})$, $K_3,K_4$ are $O((\mathcal{E}_{\mathrm{D},1}+\mathcal{E}_{\mathrm{D},2})^{M-1})$ whereas $K_{5}(k)$ and $K_{6}(k)$ are $O((\mathcal{E}_{\mathrm{D},1})^{k-1})$, $K_{7}(k)$ and $K_{8}(k)$ correspond to $O((\mathcal{E}_{\mathrm{D},1}+\mathcal{E}_{\mathrm{D},2})^{k-1})$ with $\epsilon(\rho)\in[0,1)$. 

In order to have a vanishing $P_e(\mathbf{m})$ in the first round, we set the relations of the energies as $\mathcal{E}_{C,1}=\frac{\mathcal{E}_{D,2}}{2(1-\sqrt{\lambda})^{2}}$ and $\mathcal{E}_{D,2}=(2-\mu)\mathcal{E}_{D,1}$ where $\mu$ is an arbitrary constant satisfying $\mu\in(0,2)$, so that the average energy used by the protocol 
\begin{equation}
{\mathcal{E}}\leq \mathcal{E}_{\mathrm{D},1}+\mathcal{E}_{\mathrm{C},1}P_e(\mathbf{m})+\mathcal{E}_{\mathrm{D},2}[P_e(\mathbf{m})(1-\Pr(E_{e\rightarrow c,1}|M\;\;in\;\;error))+(1-P_e(\mathbf{m}))\Pr(E_{c\rightarrow e,1})]
\end{equation}
can be made arbitrarily close to $\mathcal{E}_{\mathrm{D},1}$ guaranteed by vanishing union error probability $P_e(\mathbf{m})$ given that $k<M$. The derivation of the bound on the average energy given above can be found in Appendix \ref{subsec:appIII}. Here, $\Pr(E_{c\rightarrow e,1})$ represents the total misdetected acknowledged error probability in the first round which is equivalent to $\exp\left\{{-\frac{\lambda \mathcal {E}_{\mathrm{C},1}}{N_{0}}}\right\}$. 

For the case of high correlation, i.e. when $2^{B+1}\sqrt{1-\rho^2}<\theta$ where $\theta \sim O(1)$, we obtain the asymptotic bound with respect to $B$ on distortion as 
\begin{equation}\label{dist-highcorr_M} 
D_{high}\leq \alpha(\mathcal{E}_{\mathrm{D},1},\rho,M) \exp \left\{-\frac{\mathcal{E}_{\mathrm{D},1}(1-\mu/3)}{N_0}\right\}
+\sum_{k=1}^{M-1}\beta (k,\rho,M) \exp \left \{-\frac{(2M+3k-\mu(2M+k))\mathcal{E}_{\mathrm{D},1}}{2MN_0} \right \}
\end{equation} Because of the quantizer construction, $1-\rho^2$ is considered as in the same order of $2^{-2B}$ and consequently should be chosen to behave as $\exp \left\{-\frac{\mathcal{E}_{\mathrm{D},1}}{N_0}\right\}$. As a result we obtain the same collaboration effect as in (\ref{ineq:multd2}) albeit with a factor 2 gap in energy efficiency.  The latter may be due to simplifying steps in the outer-bound. Furthermore, we notice that condition for exploiting collaboration between the sources is based on relationship between the observation error variance ($1-\rho^2$) and the {\em aggregate energy} as opposed to the individual source energies.
The significant term is isolated in the bound given above in order to be emphasized, where 
\begin{multline}
\alpha(\rho,M)=\left(\frac{29/2+2\sqrt{3}}{\rho^2 M}+\frac{\rho^2 M+12\rho^2+12\rho \sqrt{1-\rho^2}+3(1-\rho^2)}{\rho^2 M}\right)
\\
\left((K_1'+K_2\epsilon(\rho))^{M-1}2^{-2M+2}+(K_3'+K_4\epsilon(\rho))^{M-1}2^{-4M+2}\right)
\end{multline}
which arose from the distortion terms corresponds to $K_2$, $K_4$ in (\ref{dist_M}) and $\beta$ also denotes a function of $k$, $\rho$ and $M$ which is given by 
\begin{equation} \label{beta}
\beta (k,\rho,M)=\frac{9M+16k}{(M^2 \rho^2)/3} \dbinom{M}{k}\left [(K'_5(k)+K_6(k)\epsilon(\rho))^k + (K'_7(k)+K_8(k)\epsilon(\rho))^k)\right]
\end{equation} arose from the distortion terms corresponds to $K_5$, $K_6$, $K_7$ and $K_8$ which represents the lower order terms. Note that $K'_n(k)=\theta K_n(k) $ for $n=1,3,5,7$.

\section{Practical Adaptation and Numerical Evaluation for Finite Energy \label{sec:numerical}}

In this section we consider the performance of the protocol described in the previous section with a minor adaptation allowing
finer control of the probability of going to the second round. We introduce the condition that the receiver decides to continue
with the second round only if it detects $L$ or more errors after the control phase, including the extreme case of $L=M$ which
corresponds to detecting errors from all sources.  Based on this adaptation, the bound on the distortion to be achieved at the end of the second round given above by (\ref{dist_M}) can be expanded as 
 \begin{align}
&D\leq D_q
+ D_{e,M} \left(K_{1}\frac{\sqrt{1-\rho^2}}{\rho+\sqrt{1-\rho^2}}e^{(B+1)\ln2}+K_2 \epsilon(\rho)\right)^{M-1}e^{(B-2M+2)\ln2-\frac{\mathcal{E}_{\mathrm{D},1}+2\mathcal{E}_{\mathrm{C},1}(\sqrt{\lambda}-1)^2}{2N_{0}}} \nonumber \\
&+ D_{e,M} \left(K_{3}\frac{\sqrt{1-\rho^2}}{\rho+\sqrt{1-\rho^2}}e^{(B+1)\ln2}+K_4 \epsilon(\rho)\right)^{M-1}e^{(B-4M+2)\ln2-\frac{\mathcal{E}_{\mathrm{D},1}+\mathcal{E}_{\mathrm{D},2}}{2N_{0}}}\nonumber \\
&+\sum_{k=1}^{L-1} D_{e,k} \dbinom{M}{k}  \left(K_{5}(k)\frac{\sqrt{1-\rho^2}}{\rho+\sqrt{1-\rho^2}}e^{(B+1)\ln2}+K_6(k)\epsilon(\rho)\right)^{k}e^{-\frac{k\mathcal{E}_{\mathrm{D},1}}{2MN_0}} \nonumber \\
&+\sum_{k=L}^{M-1} D_{e,k} \dbinom{M}{k}  \left(K_{5}(k)\frac{\sqrt{1-\rho^2}}{\rho+\sqrt{1-\rho^2}}e^{(B+1)\ln2}+K_6(k)\epsilon(\rho)\right)^{k}e^{-\frac{k\mathcal{E}_{\mathrm{D},1}+2(k-L+1)\mathcal{E}_{\mathrm{C},1}(\sqrt{\lambda}-1)^2}{2MN_0}} \nonumber \\
&+\sum_{k=1}^{M-1} D_{e,k} \dbinom{M}{k} \left(K_{7}(k)\frac{\sqrt{1-\rho^2}}{\rho+\sqrt{1-\rho^2}}e^{(B+1)\ln2}+K_8(k)\epsilon(\rho)\right)^{k}e^{-\frac{k(\mathcal{E}_{\mathrm{D},1}+\mathcal{E}_{\mathrm{D},2})}{2MN_0}}. \label{dist_M'}
\end{align} 
\begin{figure}[htp]
  \centering
  \includegraphics[width=0.70\linewidth]{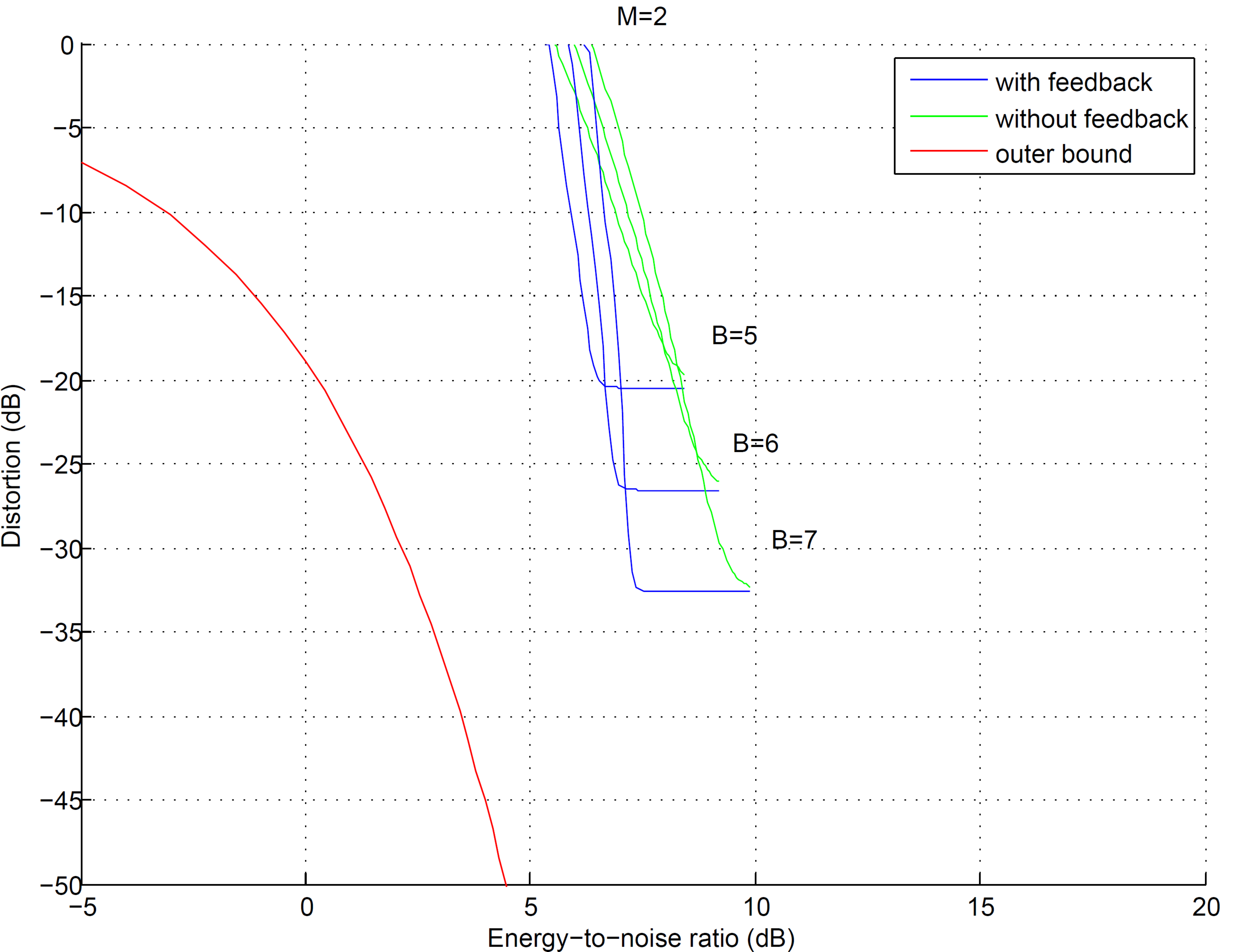}
  \caption{Numerical evaluation of \ref{dist_M'} for $M=2$ and $B$ from 5 to 7.}
	\label{fig:numeric1}
\end{figure}
\begin{figure}[htp]
  \centering
  \includegraphics[width=0.70\linewidth]{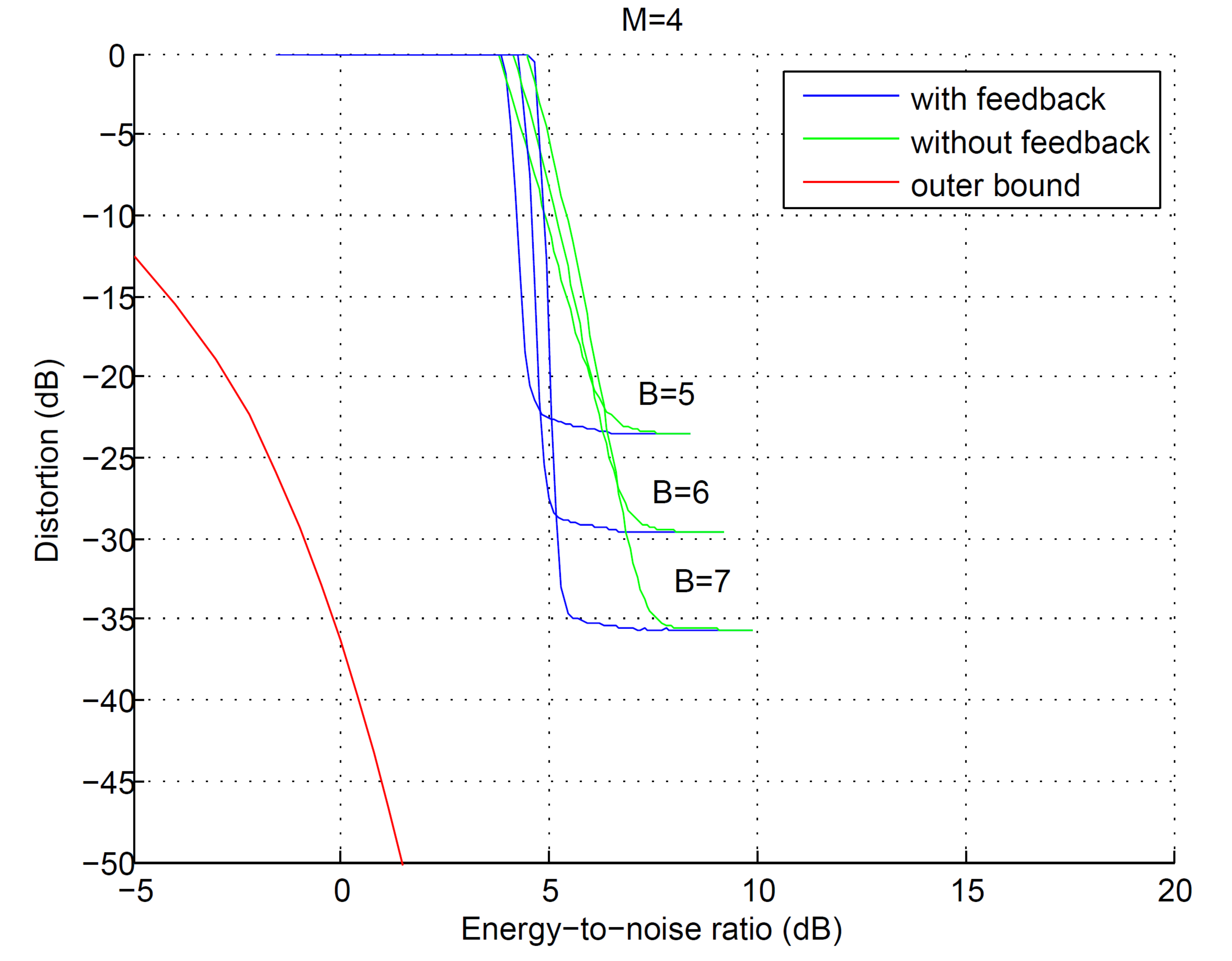}
  \caption{Numerical evaluation of \ref{dist_M'} for $M=4$ and $B$ from 5 to 7.}
	\label{fig:numeric2}
\end{figure}
\begin{figure}[htp]
  \centering
  \includegraphics[width=0.70\linewidth]{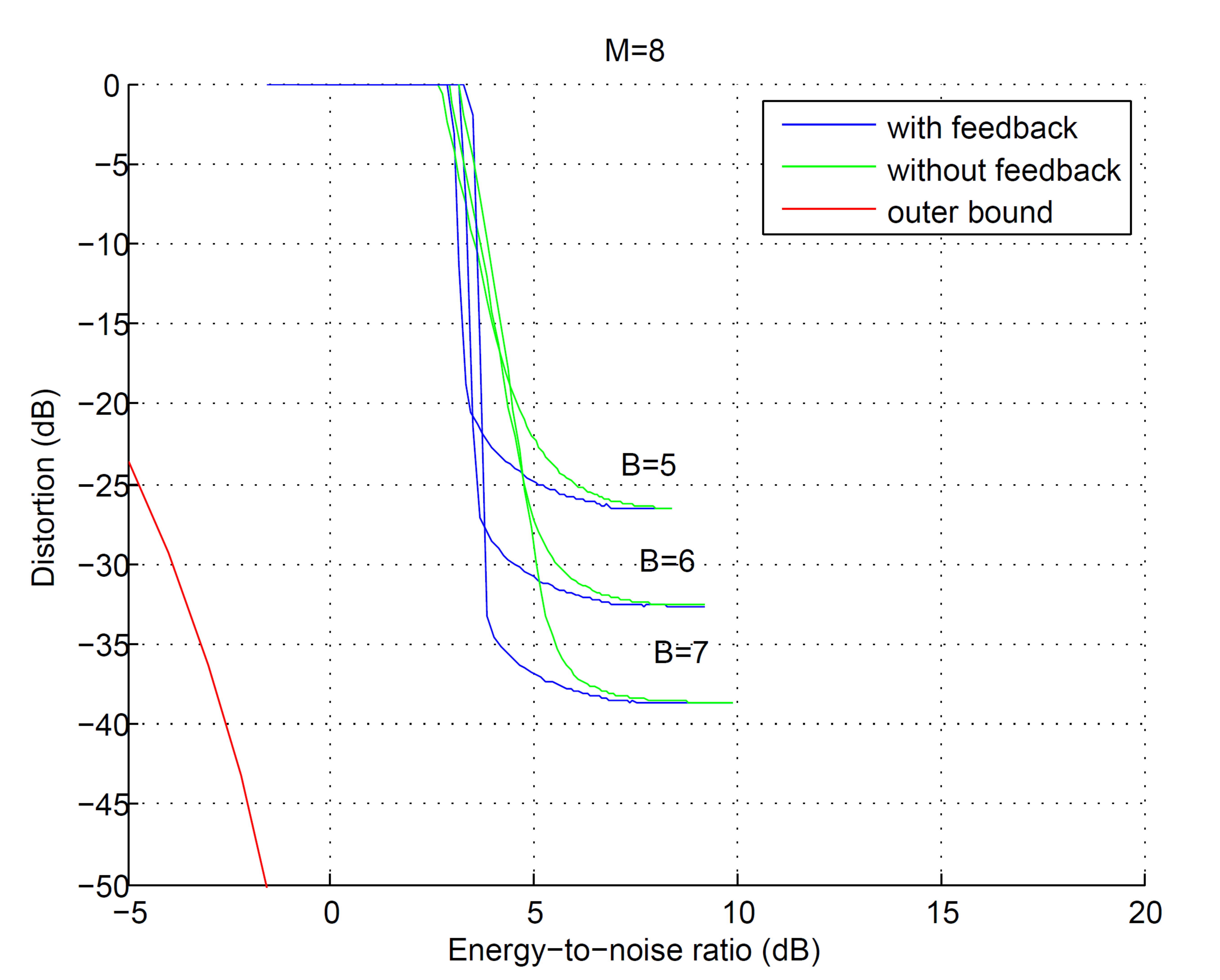}
  \caption{Numerical evaluation of \ref{dist_M'} for $M=8$ and $B$ from 5 to 7.}
	\label{fig:numeric3}
\end{figure}

The motivation for this adaptation is to bring the distortion down to $D_q$ as quickly as possible, since as long as a few sources are decoded correctly, $D$ will be proportional to $2^{-2B}$.  It is only when a significant number of sources are in error that we make use of the extra energy in the second round to bring the distortion close to $D_q$. In Figures \ref{fig:numeric1}-\ref{fig:numeric2} and \ref{fig:numeric3} we show the behaviour of the protocol for $M=2,4,8$ and $B=7$ bits compared to the lower bound on $D$ in \ref{ineq:multd} for uniform statistics on one-dimensional $\mathbf{U}$ and $\mathbf{U}_j'$. We first see that, as in the single-source and dual correlated source problems from \cite{EURECOM+4081} we do not quite approach the asymptotic gain of 4.7 dB over the case when only 1 round is used (i.e. one-shot transmission without feedback having exponent $e^{-\frac{\mathcal{E}}{3N_0}}$ as opposed to $e^{-\frac{\mathcal{E}}{N_0}}$).  We can, however, obtain somewhere between 2-3 dB for moderate energies. Secondly, we see that increasing $M$ has the effect of allowing for lower distortions around the asymptote, $D_q$, since the latter decreases linearly with $M$.  However, the energy-efficiency of the protocol does not decrease linearly with $M$ due to the non-coherent combining loss, which can even be seen when going from $M=4$ to $M=8$.  We note that the lower-bounds do improve linearly with $M$ due to the bounding step (a) in \ref{mults_C}.  It would be worthwhile to derive a lower-bound which does not remove the unknown parameter to determine if this effect is unavoidable.  Moreover, if we were to consider spatial-expansion (i.e. increasing $M$) with fixed total energy, we would have a loss in energy efficiency which increases with $M$.

\section{Conclusion \label{sec:conc}}

This paper covered an adaptation of two-way low-latency feedback protocol for minimal distortion studied in \cite{EURECOM+3798} and \cite{EURECOM+3898} to a large network scenario with multiple sources. Specifically, we have provided lower-bounds on the reconstruction
error of arbitrary multi-sensor transmission strategies which can serve in a subsequent
step to determine the optimality of particular multiple-accces and encoding strategies. To this
end, we have proposed one such collaborative strategy exploiting correlation between sensors.
Asymptotic upper-bounds on the reconstruction error have been provided for the proposed protocol.
Both the upper and lower-bounds show that collaboration can be 
achieved through energy accumulation and bring to light a
trade-off in source and channel SNR allowing it to occur. 
The practical performance of the proposed retransmission protocol was investigated through numerical evaluation of the upper-bounds in the non-asymptotic energy regime, which corresponds to using low-order quantization in the sensors. 
The performance of the protocol was improved through the introduction of a minor modification in the feedback strategy which allows the 
error-free performance to be achieved quickly. Comparisons with a one-shot transmission not exploiting feedback show that gains with one round of feedback are on the order to 2-3 dB in comparison to a feedback-less system and are often to within 5 dB from the lower-bound. It is further shown that an increase in the size of the network brings benefit in terms of performance, but that the gain in terms of energy efficiency diminishes quickly at finite energies due to a non-coherent combining loss.

Future work will consider more general distributed sensing and transmission strategies
for multi-dimensional sources aiming at energy-efficiency and low-latency protocols.  In terms of the limits studied
in this work, it would be worthwhile to consider tightening the lower-bounds (\ref{ineq:multd}) and (\ref{ineq:multdu1}) by not
implicitly assuming channel side information at the receiving end.  The latter would allow us to shed more light on the limitations of the size of the network with respect to energy efficiency, since it was seen that the lack of channel side information was a limiting factor in the performance of the proposed feedback strategy. It would also be worth studying the case of more general random-channels as for the single-source case considered in \cite{EURECOM+4081}.

\newpage
\section{Appendix} \label{app}
\subsection{Appendix I- Estimation of $U$} \label{subsec:appI}
The first expansion of (\ref{mults_D}) based on the output signals is given by
\begin{align}  \label{mults_C}
&I(\mathbf{U};\mathbf{Y}|\{\mathbf{V}_{j}\}_{S})\overset{(a)}{\leq} I(\mathbf{U};\mathbf{Y}|\{\mathbf{V}_{j}\}_{S},\Phi)\nonumber\\
&=h(\mathbf{Y}|\{\mathbf{V}_{j}\}_{S},\Phi)-h(\mathbf{Y}|\mathbf{U},\{\mathbf{V}_{j}\}_{S},\Phi)\nonumber \\
&=\sum_{i=1}^{N} h(Y_{i}|Y^{i-1},\{\mathbf{V}_{j}\}_{S},\Phi)- \sum_{i=1}^{N} h(Y_{i}|Y^{i-1},\{\mathbf{V}_{j}\}_{S},\mathbf{U},\Phi)\nonumber\\
&\leq \sum_{i=1}^{N} h(Y_{i}|Y^{i-1},\{\mathbf{V}_{j}\}_{S},\{\mathbf{X}_{j}e^{i\phi_{j}}\}_{S},\Phi)- \sum_{i=1}^{N} h(Y_{i}|Y^{i-1},\mathbf{U},\{\mathbf{X}_{j}e^{i\phi_{j}}\}_{S},\{\mathbf{X}_{j}e^{i\phi_{j}}\}_{S^c},\Phi)\nonumber\\
&=\sum_{i=1}^{N} h(Y_{i}-\sum_{j\in S}X_{i,j}e^{i\phi_{i,j}}|Y^{i-1},\{\mathbf{V}_{j}\}_{S},\{\mathbf{X}_{j}e^{i\phi_{j}}\}_{S},\Phi)\nonumber \\
&-\sum_{i=1}^{N} h(Y_{i}-\sum_{j\in S}X_{i,j}e^{i\phi_{i,j}}-\sum_{j\in S^c}X_{i,j}e^{i\phi_{i,j}}|Y^{i-1},\mathbf{U}, \{\mathbf{X}_{j}e^{i\phi_{j}}\}_{S},\{\mathbf{X}_{j}e^{i\phi_{j}}\}_{S^c},\Phi)\nonumber \\ 
&=\sum_{i=1}^{N} h(\sum_{j\in S^{c}}X_{i,j}e^{i\phi_{i,j}}+Z_{i}|Y^{i-1},\{\mathbf{V}_{j}\}_{S},\{\mathbf{X}_{j}e^{i\phi_{j}}\}_{S},\Phi)- \sum_{i=1}^{N} h(Z_{i}) \nonumber\\
&\leq \sum_{i=1}^{N} \log \left(1+\frac{\sum_{j\in S^{c}}\mathcal{E}_{i,j}}{N_0}\right) \nonumber \\
&\leq N\log \left(1+\frac{\sum_{i=1}^{N}\sum_{j\in S^{c}}\mathcal{E}_{j}}{NN_0}\right) \nonumber \\
&\leq N\log \left(1+\frac{K\sum_{j\in S^{c}}\mathcal{E}_{j}}{N N_0}\right).
\end{align} 
where $(a)$ is a result of the fact that $\mathbf{U}$ and $\Phi$ are independent.
The second expansion of $I(\mathbf{U};\mathbf{Y}|\{\mathbf{V}_{j}\}_{S})$ based on the sources, we have two different derivations for the two distribution types. 
\begin{align} \label{mults_D}
I(\mathbf{U};\mathbf{Y}|\{\mathbf{V}_{j}\}_{S})&=h(\mathbf{U}|\{\mathbf{V}_{j}\}_{S})-h(\mathbf{U}|\mathbf{Y},\{\mathbf{V}_{j}\}_{S}) \nonumber \\
&=-I(\mathbf{U};\{\mathbf{V}_{j}\}_{S})+h(\mathbf{U})-h(\mathbf{U}-\mathbf{\hat{U}}|\mathbf{Y},\{\mathbf{V}_{j}\}_{S}) \nonumber \\
&\geq-h(\{\mathbf{V}_{j}\}_{S})+h(\{\mathbf{V}_{j}\}_{S}|\mathbf{U})+h(\mathbf{U})-h(\mathbf{U}-\mathbf{\hat{U}}). 
\end{align}
For the case where $\mathbf{U}$ and whole set of $\mathbf{U}_{j}$'s are uniformly distributed, the above expansion (\ref{mults_D}) becomes
\begin{align} \label{mults_D1}
I(\mathbf{U};\mathbf{Y}|\{\mathbf{V}_{j}\}_{S})& \geq -\frac{K}{2}\log \left((2\pi e)^{|S|}(1-\rho^2)^{|S|}(1+\frac{|S|\rho^2}{1-\rho^2})\right)+ |S|K\log(2\sqrt{3(1-\rho^2)})+K\log 2\sqrt{3}-\frac{K}{2}\log(2\pi e D)\nonumber \\
&= \frac{K}{2}\log \left(\left(\frac{6}{\pi e}\right)^{|S|+1}\frac{1}{(1+\frac{|S|\rho^2}{1-\rho^2})D}\right)
\end{align} 
whereas the same expansion yields for the Gaussian case
\begin{align}
I(\mathbf{U};\mathbf{Y}|\{\mathbf{V}_{j}\}_{S})&\geq -\frac{K}{2}\log \left((2\pi e)^{|S|}(1-\rho^2)^{|S|}(1+\frac{|S|\rho^2}{1-\rho^2})\right)
+\frac{|S|K}{2}\log (1-\rho^2)2\pi e+ \frac{K}{2}\log (2\pi e)-\frac{K}{2}\log (2\pi e D) \nonumber \\
&=\frac{K}{2}\log \left(\frac{1}{(1+\frac{|S|\rho^2}{1-\rho^2})D}\right)
\end{align}
where $|S|$ denotes the size of the set $\mathbf{V}_{j}$ and using the following bound on entropy $h(\mathbf{U}-\mathbf{\hat{U}})$
\begin{align}
h(\mathbf{U}-\mathbf{\hat{U}})&\leq \sum_{j=1}^{K}h(U_{j}-\hat{U}_{j})\nonumber \\
&\leq \frac{K}{2}\log \left(\frac{2\pi e}{K}\sum_{j=1}^{K}\mathbb{E}[(U_{j}-\hat{U}_{j})^2]\right) \nonumber \\
&\leq K\log \left(\sqrt{2\pi e D}\right).
\end{align} 

\subsection{Appendix II- Estimation of the set of $V_j$'s} \label{subsec:appII}
\begin{align} \label{mults_E}
&I(\mathbf{V}_{j};\mathbf{Y}|\{\mathbf{V}_{l}\}_{S})\overset{(a)}{\leq}I(\mathbf{V}_{j};\mathbf{Y}|\{\mathbf{V}_{l}\}_{S},\Phi)\nonumber \\
&=h(\mathbf{Y}|\{\mathbf{V}_{l}\}_{S},\Phi)-h(\mathbf{Y}|\mathbf{V}_{j},\{\mathbf{V}_{l}\}_{S},\Phi)\nonumber \\
&=\sum_{i=1}^{N} h(Y_{i}|Y^{i-1},\{\mathbf{V}_{l}\}_{S},\Phi)- \sum_{i=1}^{N} h(Y_{i}|Y^{i-1},\mathbf{V}_{j},\{\mathbf{V}_{l}\}_{S},\Phi) \nonumber \\
&\leq \sum_{i=1}^{N} h(Y_{i}|Y^{i-1},\{\mathbf{V}_{l}\}_{S},\{\mathbf{X}_{j}e^{i\phi_{j}}\}_{S},\Phi)- \sum_{i=1}^{N} h(Y_{i}|Y^{i-1},\mathbf{V}_{j},\{\mathbf{V}_{l}\}_{S},\{\mathbf{X}_{j}e^{i\phi_{j}}\}_{S^c},\{\mathbf{X}_{j}e^{i\phi_{j}}\}_{S},\Phi) \nonumber \\
&=\sum_{i=1}^{N} h(Y_{i}-\sum_{j\in S}X_{i,j}e^{i\phi_{i,j}}|Y^{i-1},\{\mathbf{V}_{l}\}_{S},\{\mathbf{X}_{j}e^{i\phi_{j}}\}_{S},\Phi)\nonumber \\
&-\sum_{i=1}^{N} h(Y_{i}-\sum_{j\in S^c}X_{i,j}e^{i\phi_{i,j}}-\sum_{j\in S}X_{i,j}e^{i\phi_{i,j}}|Y^{i-1},\{\mathbf{V}_{j}\}_{S^c},\{\mathbf{V}_{l}\}_{S},\{\mathbf{X}_{j}e^{i\phi_{j}}\}_{S^c},\{\mathbf{X}_{j}e^{i\phi_{j}}\}_{S},\Phi) \nonumber \\
&\leq \sum_{i=1}^{N} h(\sum_{j\in S^c}X_{i,j}e^{i\phi_{i,j}}+Z_{i}|Y^{i-1},\{\mathbf{V}_{l}\}_{S},\{\mathbf{X}_{j}e^{i\phi_{j}}\}_{S},\Phi)- \sum_{i=1}^{N} h(Z_{i}) \nonumber \\
&\leq \sum_{i=1}^{N} \log \left(1+\frac{\sum_{j\in S^c}\mathcal{E}_{i,j}}{NN_0}\right) \nonumber \\
&\leq N\log \left(1+\frac{\sum_{i=1}^{N}\sum_{j\in S^c}\mathcal{E}_{i,j}}{NN_0}\right) \nonumber \\
&\leq N\log \left(1+\frac{K\sum_{j\in S^c}\mathcal{E}_{j}}{N N_0}\right).
\end{align} where $(a)$ is a result of the fact that $\mathbf{V}_{j}$ and $\Phi$ are independent. The second expansion is obtained for the Gaussian sources as
\begin{align}
I(\mathbf{V}_{j};\mathbf{Y}|\{\mathbf{V}_{l}\}_{S})&=h(\mathbf{V}_{j}|\{\mathbf{V}_{l}\}_{S})-h(\mathbf{V}_{j}|\mathbf{Y},\{\mathbf{V}_{l}\}_{S})\nonumber \\
&=-I(\mathbf{V}_{j};\{\mathbf{V}_{l}\}_{S})+h(\mathbf{V}_{j})-h(\mathbf{V}_{j}-\mathbf{\hat{V}}_{j}|\mathbf{Y},\{\mathbf{V}_{l}\}_{S}) \nonumber \\
&\geq -h(\{\mathbf{V}_{l}\}_{S})+h(\{\mathbf{V}_{l}\}_{S}|\mathbf{V}_{j})+h(\mathbf{V}_{j})-h(\mathbf{V}_{j}-\mathbf{\hat{V}}_{j})\nonumber \\
&=-\frac{K}{2}\log \left((2\pi e)^{|S|}(1-\rho^2)^{|S|}\left(1+\frac{|S|\rho^2}{1-\rho^2}\right)\right)+\frac{K}{2}\log \left((2\pi e)^{|S|}(1-\rho^2)^{|S|}(2-\rho^2+|S|)\right)\nonumber \\
&+\frac{K}{2}\log(2\pi e)-\frac{K}{2}\log (2\pi e D)\nonumber \\
&= \frac{K}{2}\log \left(\frac{1}{D}\frac{(1-\rho^2)(2-\rho^2+|S|)}{1+(|S|-1)\rho^2}\right)  \label{mults_F}
\end{align} whereas for the uniform case, same mutual information yields
\begin{align} \label{mults_G}
&I(\mathbf{V}_{j};\mathbf{Y}|\{\mathbf{V}_{l}\}_{S})=h(\mathbf{V}_{j}|\{\mathbf{V}_{l}\}_{S})-h(\mathbf{V}_{j}|\mathbf{Y},\{\mathbf{V}_{l}\}_{S})\nonumber \\
&=-I(\mathbf{V}_{j};\{\mathbf{V}_{l}\}_{S})+h(\mathbf{V}_{j})-h(\mathbf{V}_{j}-\mathbf{\hat{V}}_{j}|\mathbf{Y},\{\mathbf{V}_{l}\}_{S}) \nonumber \\
&\overset{(a)}{\geq}-h(\{\mathbf{V}_{l}\}_{S})+h(\{\mathbf{V}_{l}\}_{S}|\mathbf{V}_{j})+h(\mathbf{V}_{j})-h(\mathbf{V}_{j}-\mathbf{\hat{V}}_{j})\nonumber \\
&\geq -h(\{\mathbf{V}_{l}\}_{S})+\frac{K}{2} \log \left(2^{\frac{2}{K}h(\rho \mathbf{U}\mathbf{1}|\mathbf{V}_{j})}+2^{\frac{2}{K}h(\{\sqrt{1-\rho^2}\mathbf{U'}_{j}\}_{S}|\mathbf{V}_{j})}\right)+\frac{K}{2}\log \left(2^{\frac{K}{2}h(\rho \mathbf{U})}
+2^{\frac{K}{2}h(\sqrt{1-\rho^2}\mathbf{U'}_{j})}\right)\nonumber \\
& \quad \quad \quad -h(\mathbf{V}_{j}-\mathbf{\hat{V}}_{j})\nonumber \\
&= -\frac{K}{2}\log\left((2\pi e)^{|S|}(1-\rho^2)^{|S|}(1+\frac{|S|\rho^2}{1-\rho^2})\right)+\frac{K}{2}\log \left(2^{\frac{2}{K}h(\rho \mathbf{U}\mathbf{1}|\mathbf{V}_{j})}+2^{\frac{2}{K}h(\{\sqrt{1-\rho^2}\mathbf{U'}_{j}\}_{S}|\mathbf{V}_{j})}\right) \nonumber \\
& \quad \quad \quad +\frac{K}{2}\log(12)-\frac{K}{2}\log(2\pi e D)
\end{align}
with
\begin{align}  \label{mults_G1}
h(\{\sqrt{1-\rho^2}\mathbf{U'}_{j}\}_{S}|\mathbf{V}_{j})&\overset{(a)}{=}h(\{\sqrt{1-\rho^2}\mathbf{U'}_{j}\}_{S})\nonumber \\
&=\frac{K}{2}\log (12(1-\rho^2))^{|S|}
\end{align}
(a) is caused by the independence between $\mathbf{U}_{j}'$ and $\mathbf{V}_{j}$. For the completion of (\ref{mults_G}) last expression to be derived is the conditional entropy $h(\rho \mathbf{U}\mathbf{1}|\mathbf{V}_{j})$ which has the expansion given below.
\begin{align}  \label{mults_G2}
h(\rho \mathbf{U}\mathbf{1}|\mathbf{V}_{j})&=h(\rho \mathbf{U}\mathbf{1},\rho \mathbf{U}+\sqrt{1-\rho^2}\mathbf{U'}_{j})-h(\rho \mathbf{U}+\sqrt{1-\rho^2}\mathbf{U'}_{j})\nonumber \\
&\geq h(\rho \mathbf{U}\mathbf{1})+h(\rho \mathbf{U}+\sqrt{1-\rho^2}\mathbf{U'}_{j}|\rho \mathbf{U}\mathbf{1})-\frac{K}{2}\log(2\pi e)\nonumber \\
&=\frac{K}{2}\log(12\rho^2)+\frac{K}{2}\log(12(1-\rho^2))-\frac{K}{2}\log(2\pi e)\nonumber \\
&=\frac{K}{2}\log \left(\frac{72\rho^2(1-\rho^2)}{\pi e}\right)
\end{align} Substituting (\ref{mults_G1}) and (\ref{mults_G2}) into (\ref{mults_G}) yields 
\begin{equation}
I(\mathbf{V}_{j};\mathbf{Y}|\{\mathbf{V}_{l}\}_{S})\geq \frac{K}{2} \log \left(\frac{12\left(\frac{72 \rho^2(1-\rho^2)}{\pi e}+12^{|S|(1-\rho^2)^{|S|}}\right)}{D(2 \pi e)^{|S|+1}(1-\rho^2)^{|S|-1}(1+(|S|-1)\rho^2)}\right).
\end{equation}

\subsection{Appendix III- Bound on product Distortion $\prod_{j=1}^{M}D_j$} \label{subsec:appIV}
 For the first expansion based on the output signals, we have
\begin{align} \label{mults_J1}
&I(\{\mathbf{V}_{j}\};\mathbf{Y})\overset{(a)}{\leq}I(\{\mathbf{V}_{j}\};\mathbf{Y}|\Phi)\nonumber \\
&=h(\mathbf{Y}|\Phi)-h(\mathbf{Y}|\{\mathbf{V}_{j}\},\Phi)\nonumber \\
&=\sum_{i=1}^{N} h(Y_{i}|Y^{i-1},\Phi)- \sum_{i=1}^{N} h(Y_{i}|Y^{i-1},\{\mathbf{V}_{j}\},\Phi) \nonumber \\
&=\sum_{i=1}^{N} h(Y_{i}|Y^{i-1},\Phi)- \sum_{i=1}^{N} h(Y_{i}|Y^{i-1},\{\mathbf{V}_{j}\},\{\mathbf{X}_{j}e^{i\phi_{j}}\},\Phi) \nonumber \\
&=\sum_{i=1}^{N} h(\sum_{j=1}^{M} X_{i,j}e^{i\phi_{i,j}}+Z_i|Y^{i-1},\Phi)-\sum_{i=1}^{N} h(Y_{i}-\sum_{j=1}^{M}X_{i,j}e^{i\phi_{i,j}}|Y^{i-1},\{\mathbf{V}_{j}\},\{\mathbf{X}_{j}e^{i\phi_{j}}\},\Phi) \nonumber \\
&= \sum_{i=1}^{N} h(\sum_{j=1}^{M}X_{i,j}e^{i\phi_{i,j}}+Z_{i}|Y^{i-1},\Phi)- \sum_{i=1}^{N} h(Z_{i}) \nonumber \\
&\leq \sum_{i=1}^{N} \log \left(1+\frac{\sum_{j=1}^{M}\mathcal{E}_{i,j}}{NN_0}\right) \nonumber \\
&\leq N\log \left(1+\frac{\sum_{i=1}^{N}\sum_{j=1}^{M}\mathcal{E}_{i,j}}{NN_0}\right) \nonumber \\
&\leq N\log \left(1+\frac{KM\mathcal{E}}{N N_0}\right).
\end{align} where $(a)$ is a result of the fact that $\mathbf{V}_{j}$ and $\Phi$ are independent. 
On the other hand, the second expansion of $I(\{\mathbf{V}_{j}\};\mathbf{Y})$ is derived for normally distributed $\mathbf{V}_{j}$'s as
\begin{align}
I(\{\mathbf{V}_{j}\};\mathbf{Y})&=h(\{\mathbf{V}_{j}\})-h(\{\mathbf{V}_{j}\}|\mathbf{Y})\nonumber \\
&=h(\{\mathbf{V}_{j}\})-h(\{\mathbf{V}_{j}-\mathbf{\hat{V}}_{j}\}|\mathbf{Y}) \nonumber \\
&\geq h(\{\mathbf{V}_{j}\})-h(\{\mathbf{V}_{j}-\mathbf{\hat{V}}_{j}\})\nonumber \\
&= \frac{K}{2}\log \left((1-\rho^2)^M (2\pi e)^{M}(1+\frac{M\rho^2}{1-\rho^2})\right)-\frac{K}{2}\log \left((2\pi e)^M \prod_{j=1}^{M}D_j \right)\nonumber \\
&= \frac{K}{2}\log \left(\frac{(1-\rho^2)^{M}(1+\frac{M\rho^2}{1-\rho^2})}{\prod_{j=1}^{M}D_j}\right). \label{mults_K1}
\end{align} which yields
\begin{align}
I(\{\mathbf{V}_{j}\};\mathbf{Y})&=h(\{\mathbf{V}_{j}\})-h(\{\mathbf{V}_{j}\}|\mathbf{Y})\nonumber \\
&=h(\{\mathbf{V}_{j}\})-h(\{\mathbf{V}_{j}-\mathbf{\hat{V}}_{j}\}|\mathbf{Y}) \nonumber \\
&\geq h(\{\mathbf{V}_{j}\})-h(\{\mathbf{V}_{j}-\mathbf{\hat{V}}_{j}\})\nonumber \\
&=h(\{\rho \mathbf{U}\mathbf{1}+\sqrt{1-\rho^2}\mathbf{U'}_{j}\})-h(\{\mathbf{V}_{j}-\mathbf{\hat{V}}_{j}\})\nonumber \\
&\geq \frac{KM}{2}\log \left(2^{\frac{2}{MK}h(\rho \mathbf{U}\mathbf{1})}+2^{\frac{2}{MK}h(\{\sqrt{1-\rho^2}\mathbf{U'}_{j}\})}\right)-\frac{K}{2}\log \left((2\pi e)^M \prod_{j=1}^{M}D_j \right)\nonumber \\
&=\frac{K}{2} \log \left(\left(\frac{(12\rho^2)^{1/M}+12(1-\rho^2)}{2\pi e}\right)^M \frac{1}{\prod_{j=1}^{M}D_j}\right)
\end{align} for the uniform case.

\subsection{Appendix IV- Derivation of the Bounds on Distortion and the Average Energy of the Protocol} \label{subsec:appIII}
\begin{align} \label{gendist_M'}
&D(\mathcal{E},N_{0},2,\lambda) =D_q(1-P_e)+ \sum_{k=1}^{M-1} D_{e,k}P_{e,k}^{(1)}\Pr(E_{e\rightarrow c,1}|k\;\;in\;\;error)+D_{e,M} P_{e,M}^{(1)}\Pr(E_{e\rightarrow c,1}|M\;\;in\;\;error) \nonumber \\
&+\Pr(E_1)\left(\sum_{k=1}^{M-1} D_{e,k}P_{e,k}^{(2)}(E_1)(1-\Pr(E_{e\rightarrow c,1}|k\;\;in\;\;error))+D_{e,M} P_{e,M}^{(2)}(E_1)(1-\Pr(E_{e\rightarrow c,1}|M\;\;in\;\;error))\right)\nonumber \\
&+\Pr(E_1^c)\Pr(E_{c\rightarrow e,1})\left(\sum_{k=1}^{M-1} D_{e,k}P_{e,k}^{(2)}(E_1^c)+ D_{e,M}P_{e,M}^{(2)}(E_1^c)\right) \nonumber \\
&\overset{(a)}{\leq} D_q + \sum_{k=1}^{M-1} D_{e,k}P_{e,k}^{(1)}\Pr(E_{e\rightarrow c,1}|k\;\;in\;\;error)+D_{e,M} P_{e,M}^{(1)}\Pr(E_{e\rightarrow c,1}|M\;\;in\;\;error)\nonumber \\
&+ \left(\Pr(E_1)\sum_{k=1}^{M-1} D_{e,k}P_{e,k}^{(2)}(E_1) + \Pr(E_1^c)\sum_{k=1}^{M-1} D_{e,k}P_{e,k}^{(2)}(E_1^c)\right)+ \left(\Pr(E_1)D_{e,M} P_{e,M}^{(2)}(E_1)+\Pr(E_1^c)D_{e,M}P_{e,M}^{(2)}(E_1^c)\right)\nonumber \\
&= D_q+\sum_{k=1}^{M-1} D_{e,k}P_{e,k}^{(1)}\Pr(E_{e\rightarrow c,1}|k\;\;in\;\;error)+D_{e,M} P_{e,M}^{(1)}\Pr(E_{e\rightarrow c,1}|M\;\;in\;\;error) \nonumber \\
&+\sum_{k=1}^{M-1} D_{e,k}P_{e,k}^{(2)}+ D_{e,M}P_{e,M}^{(2)} \nonumber \\
&\leq D_q+\sum_{k=1}^{M-1} D_{e,k}P_{e,k}+D_{e,M} P_{e,M} 
\end{align}
where $P_e$ is the probability of at least one source being in error upon completion of the protocol, $\Pr(E_1)$ is the probability of at least one source being in error after the first round of the protocol (event $E_1$) whereas its complement is denoted by $\Pr(E_1^c)$. 
In step (a) the probabilities of error to be detected given $k$ and $M$ sources are in error and the probability of mis-detection $\left(\Pr(E_{c\rightarrow e,1})\right)$ are bounded by 1. The probability of not making an error at the end of the second round, i.e. $(1-P_e)$, is also upper bounded by 1 in the same step.
$P_{e,l}^{(1)}$ is the notation for $\Pr(\text{l in error after round 1})$ whereas $P_{e,l}^{(2)}$ denotes $\Pr(\text{l in error after round 2})$ for $l=1,2,...,M$.

The average energy used by protocol can be derived and bounded as
\begin{align}
{\mathcal{E}}&= \mathcal{E}_{\mathrm{D},1}+\sum_{k=1}^{M-1} \frac{k\mathcal{E}_{\mathrm{C},1}}{M}P_{e,k}^{(1)}+\mathcal{E}_{\mathrm{C},1}P_{e,M}^{(1)} \nonumber \\
&+\mathcal{E}_{\mathrm{D},2}\left[\sum_{k=1}^{M-1}P_{e,k}^{(1)}\left(1-\Pr(E_{e \rightarrow c,1}|k\;\;in\;\;error)\right)
+P_{e,M}^{(1)}\left(1-\Pr(E_{e \rightarrow c,1}|M\;\;in\;\;error)\right)\right]\nonumber \\
&+\mathcal{E}_{\mathrm{D},2}\left[\left(1-\left(\sum_{k=1}^{M-1}P_{e,k}^{(1)}+P_{e,M}^{(1)}\right)\right)\Pr(E_{c\rightarrow e,1})\right]\nonumber \\
&\leq \mathcal{E}_{\mathrm{D},1}+\mathcal{E}_{\mathrm{C},1}P_e(\mathbf{m})+\mathcal{E}_{\mathrm{D},2}[P_e(\mathbf{m})(1-\Pr(E_{e\rightarrow c,1}|M\;\;in\;\;error))+(1-P_e(\mathbf{m}))\Pr(E_{c\rightarrow e,1})]
\end{align}

\bibliography{paper_5}
\bibliographystyle{IEEEtran}


\end{document}